\documentclass[11pt] {article}

\usepackage{fullpage,amsmath,amsfonts,comment,amsthm}

\newcommand{\brac}[1]{\left[{#1}\right]}
\newcommand{\norm}[1]{\lVert{#1}\rVert}
\newcommand{\parenth}[1]{\left({#1}\right)}
\newcommand{\set}[1]{\left\{ {#1} \right\} }
\newcommand{\inner}[1]{\langle {#1} \rangle }
\newcommand{\abs}[1]{\lvert{#1}\rvert}
\newcommand{\ve}{\text{ and }}

\DeclareMathOperator{\spn}{span}

\newcommand{\R}{\mathbb{R}}

\newcommand{\remove}[1]{}

\def\epsilon{\varepsilon}
\def\phi{\varphi}

\DeclareMathOperator{\E}{ {\mathbb E} }
\DeclareMathOperator{\pr}{ {\mathbb P} }

\def\R{{\mathbb R}}

\newtheorem{theorem}{Theorem}

\newtheorem{corollary}[theorem]{Corollary}
\newtheorem{claim}[theorem]{Claim}

\newtheorem{lemma}[theorem]{Lemma}
\newtheorem{definition}[theorem]{Definition}

\usepackage[dvipsnames]{xcolor}

\newcommand{\Authornote}[2]{{\sffamily\small\color{red}{[#1: #2]}}}
\newcommand{\Authorcomment}[2]{{\sffamily\small\color{gray}{[#1: #2]}}}
\newcommand{\Authorstartcomment}[1]{\sffamily\small\color{gray}[#1: }

\newcommand{\Authorfnote}[2]{\footnote{\color{red}{#1: #2}}}
\newcommand{\Authorfixme}[1]{\Authornote{#1}{\textbf{??}}}
\newcommand{\Authormarginmark}[1]{\marginpar{\textcolor{red}{\fbox{\Large #1:!}}}}
\remove{
\newcommand{\Authornote}[2]{}
\newcommand{\Authorcomment}[2]{}
\newcommand{\Authorstartcomment}[1]{}

\newcommand{\Authorfnote}[2]{}
\newcommand{\Authorfixme}[1]{}
\newcommand{\Authormarginmark}[1]{}
}

\begin {document}

\title{\bf Approximation of non-boolean 2CSP}
\author{Guy Kindler\footnote{{\tt gkindler@cs.huji.ac.il.} School of Computer Science and Engineering, Hebrew University. Part of this work was done while visiting the Simons Institute. 
Supported by Israeli Science Fund
    Grant No. 1692/13, and Binational Science Foundation Grant No. 2012220.} \and Alexandra Kolla\footnote{{\tt alexkolla@gmail.com.} Computer Science Department, UIUC. Part of this work was done while visiting the Simons Institute. This material is based upon work supported by the National Science Foundation under Grant No. 1423452.} \and Luca Trevisan\footnote{{\tt luca@berkeley.edu.} EECS Department and Simons Institute, U.C. Berkeley. This material is based upon work supported by the National Science Foundation under Grant No. 1216642 and by the US-Israel Binational Science Foundation under Grant No. 2010451.}}
\date{}

\maketitle

\begin{abstract}
We develop a polynomial time $\Omega\left ( \frac 1R \log R \right)$ approximate algorithm for Max 2CSP-$R$, the problem where we are given a collection of constraints, each involving two variables, where each variable ranges over a set of size $R$, and we want to find an assignment to the variables that maximizes the number of satisfied constraints. Assuming the Unique
Games Conjecture, this is the best possible approximation up to constant factors.

Previously, a $1/R$-approximate algorithm was known, based on linear programming. Our algorithm is based on semidefinite programming (SDP) and on a novel rounding technique. The SDP that we use has an almost-matching integrality gap.
\end{abstract}

\setcounter{page}{0}

\newpage
\setcounter{page}{1}

\section{Introduction}

We study Max 2CSP-$R$ problems, 
that is constraint satisfaction problems with two variables per constraint, where each variable can
take values from a finite set $\Sigma$ of size $R$, and the goal is to find an assignment to the variables
that maximizes the number of satisfied constraints. For example the following is an instance of Max 2CSP-3
where $\Sigma = \{ 0,1,2 \}$, with three variables and five constraints

\[ \begin{array}{l}
X_1 \neq X_3\\
X_1 + X_2 \equiv 1 \bmod 3\\
(X_2 =0) \vee (X_1=1)\\
X_2 = X_3\\
X_1 + X_3 \equiv 2 \bmod 3
\end{array} \]

The above instance is satisfiable, as witnessed by the assignment $(X_1,X_2,X_3) = (0,1,1)$.

We devise a polynomial time $\Omega \left( \frac { \log R} {R} \right)$ approximate algorithm based on semidefinite programming.
Previously, the best approximation for this problem was $1/R$, based on linear programming \cite{STX98}.

The performance of our algorithms is best possible up to multiplicative constants and assuming the Unique Games Conjecture.
Max 2LIN-$R$ is the special case of Max 2CSP-$R$ in which every constraint is of the form $X_i - X_j = b_{i,j} \pmod R$; 
The results of Khot et al. \cite{KKMO07} imply that there are constants $c_1,c_2$ such that it is UG-hard to distinguish Max 2LIN-$R$ instances in which at most a $c_1/R$ fraction of constraints are satisfiable from instances in which at least a $c_2/\log R$ fraction of constraints are. We discuss in the Appendix how to derive the above hardness as a corollary of results in \cite{KKMO07}.

Chan \cite{C13} shows that there is a constant $c_3$ such that for every $\epsilon>0$ and every sufficiently large $R$, it is NP-hard to distinguish $(1-\epsilon)$-satisfiable instances of Max 2CSP-$R$ from instances in which at most a $c_3 \log R / \sqrt R$ fraction of constraints are satisfiable. Chan's hardness result is the strongest known result for nearly satisfiable instances of Max 2CSP-$R$, and it remains an open question whether it is possible to achieve an approximation ratio significantly better than $\log R / R$ on such instances.

As we discuss in the Appendix, it follows from results of Khot and Vishnoi \cite{KV05} that the Semidefinite Programming relaxation that we use has a $\Omega(R/(\log R)^2)$ integrality gap, which almost matches our algorithm.

\subsection{Other related previous work}

Previous algorithmic and complexity-theoretic work on Max 2CSP-$R$ has been motived by the {\em Sliding Scale Conjecture}, the {\em Unique Games Conjecture} and the question of {\em Approximation Resistance}. We summarize the main known results below.

\subsubsection*{The Sliding Scale Conjecture}

The approximability of Max 2CSP-$|\Sigma|$ is closely related to the power of PCP systems where the verifier makes two
queries and each query is answered by an element of  $\Sigma$. In particular, if we denote by $PCP_{c,s} [ O(\log n), 2, R ]$ the
set of languages that admit a 2-query PCP system with completeness $c$, soundness $s$, randomness $O(\log n)$, and answers coming from a set of size $R$, then we have that Max 2CSP-$R$ has a polynomial time $r$-approximate algorithm if and only if $PCP_{c,s} [ O(\log n), 2, R] = P$ for all $c,s$ such that $s/c < r$.

(A strong form of) the  {\em Sliding Scale Conjecture}, formulated by Bellare et al. \cite{BGLR93} is that Max 2CSP-$R$ is NP-hard to approximate within a factor $R^{\Omega(1)}$ even when $R$ is polynomially related to the number of variables and constraints.

Chan's result on 2-query PCPs \cite{C13} implies that, for constant $R$, the hardness of approximation grows with $R$ as  $\Omega( \frac  {\sqrt R} {\log R} )$. 
The reduction establishing the hardness of approximation runs in time $2^{O(R)} \cdot n^{O(\log R)}$, and so, under the assumption that NP does not have subexponential time algorithms, it establishes a $\Omega (R^{1/2} /\log R)$ hardness also for super-constant $R$, up to $R$ being roughly logarithmic in the number of variables. Raz's parallel repetition theorem establishes the NP-hardness of approximating Max 2CSP-$R$ within a factor of $R^{\epsilon}$, for an absolute constant $\epsilon >0$, and the reduction runs in time $n^{O(\log R)}$, establishing a hardness results for larger values of $R$, up to $2^{\log^{1-\delta} n}$ for every $\delta>0$.  

\subsubsection*{Approximation resistance}

H\aa stad \cite{H08} proved that Max 2CSP-$R$ problems are never {\em approximation resistant}. Given an instance of Max 2CSP-$R$ in which every constraint has $R^2 - t$ satisfying assignments, H\aa stad's algorithm gives a $ 1- \frac {t}{R^2}  + \Omega\left (  \frac t {R^4 \log R}  \right)$ factor approximation, which is better than the worst-case approximation ratio provided by picking a random assignment. Note that, for instances whose constraints have few satisfying assignments, H\aa stad's algorithm does not improve the $1/R$-approximation factor mentioned above.

\subsubsection*{Unique games}

A Unique Game is an instance of Max 2CSP-$R$ in which every constraint is of the form $x = f(y)$, where $x$ and $y$ are variables and $f: \Sigma \to \Sigma$ is a bijection. Feige and Reichman \cite{FR04} prove that there is a constant $\delta>0$ such that it is NP-hard to 
$1/R^\delta$-approximate Unique Games. Khot's Unique Games Conjecture \cite{K02} concerns the approximability of Unique Games in nearly-satisfiable instances, and there has been considerable algorithmic work devoted to testing the limits of the conjecture. Charikar, Makarychev and Makarychev \cite{CMM06} give a polynomial time algorithm that,
given a $(1-\epsilon)$-satisfiable instance of Unique Games satisfies a

\[ \Omega \left( \min \left \{ 1, \frac 1 {\sqrt {\epsilon \log R}} \right\} \cdot (1-\epsilon)^2 \cdot \left( \frac R {\sqrt {\log R}} \right) ^{\epsilon/(2-\epsilon) } \right) \]

\noindent fraction of constraints. Using the fact that a random assignment satisfies a $1/R$ fraction of constraints, one can see that the best between the outcome of the algorithm of Charikar et al. and a random assignment achieves a $\Omega(\log R/ (R \log\log R))$ approximation for Unique Games. The rounding technique developed by Charikar et al. is tailored to Unique Games problems and, while it can be adapted to ``$k$-to-1'' constraints, it seems necessary to adopt significantly different ideas when working with general 2CSP problems. 

\subsection{Our Techniques}

Our polynomial time $\Omega \left( \frac 1R \log R \right)$ approximate algorithm for Max 2CSP-$R$ is based on semidefinite programming.  Before describing our semidefinite programming relaxation, we note that we can assume without loss of generality that all constraints are of the form $(X_i = a) \wedge (X_j = b)$ and have exactly one satisfying assignment.\footnote{The reduction
from general instances to such instances is standard and simple: given an arbitrary instance $I$ of Max 2CSP-$R$ we can replace each constraint of $I$ the form $f(X_i,X_j)=1$ having $s$ satisfying assignments $\{ (a_1,b_1), \ldots, (a_s,b_s) \}$ with the $s$ constraints $(X_i = a_1) \wedge (X_j = b_1)$, $\ldots$, $(X_i = a_s) \wedge (X_j=b_s)$; if we call $I'$ the new collection of constraints generated by this substitution we see that every assignment satisfies the same number of constraints in $I$ as in $I'$.}

Given an instance of Max 2CSP-$R$ with variables ranging over a set $\Sigma$ of size $R$, in the semidefinite programming relaxation, we have a vector $x_{i,a}$ for each variable $X_i$ and for each value $a\in \Sigma$, with the intended meaning that $x_{i,a}$ is of unit length if $x_i = a$ and $x_{i,a}$ is of length zero otherwise. This is a standard approach when formulating
SDP relaxations of problems in which we have to model non-boolean assignments to variables, such as graph coloring problems and unique games.

The relaxation is

\begin{equation} \label{sdp} \begin{array}{cll}
  \max &\displaystyle  \sum_{C= (X_i = a) \wedge (X_j = b) } \langle x_{i,a} , x_{i,b} \rangle\\
  \mbox{s.t.}\\
  & \sum_a \norm{x_{i,a}}^2 = 1 & \forall i\\
  & \langle x_{i,a},x_{i,b}\rangle = 0& \forall i, \forall a\neq b\\
    & \langle x_{i,a},x_{j,b}\rangle \geq 0& \forall i,j \forall a, b
\end{array} \end{equation}

To see that the above SDP formulation is a relaxation, consider an assignment $(a_1,\ldots,a_n)$ to the variables
$(X_1,\ldots,X_n)$ of the constraint satisfaction problem; we can turn it into a feasible solution for the SDP by fixing a unit
vector $x_0$ and then letting $x_{i,a_i} = x_0$ and $x_{i,a} = {\bf 0}$ for every $i$ and for every $a\neq a_i$. We can see that such a solution is feasible and that the objective function is equal to the number of constraints satisfied by $(a_1,\ldots,a_n)$.

Before discussing our rounding algorithm, let us first see a simple rounding that gives a $1/R$ approximation. Given
a solution of the SDP, we pick a random assignment for the variables $X_i$, by assigning each variable $X_i$ independently, and by giving to $X_i$ the value $a$ with probability proportional to $|| x_{i,a}||$. That is, we create a random assignment $Z_1,\ldots,Z_n$ where the $Z_i$ are independent and
\[ \Pr [ Z_i = a ] := \frac { ||x_{i,a}|| }{\sum_{b\in \Sigma} ||x _{i,b} || } \]
Note that, by Cauchy-Schwarz we have
\[ \sum_{b\in \Sigma} || x_{i,b}|| \leq \sqrt {R} \cdot \sqrt{\sum_{b\in \Sigma} || x_{i,b}||^2} = \sqrt R \] 
Then the probability that a constraint $(X_i = a) \wedge (X_j = b)$ is satisfied is
\[ \Pr [ (Z_i = a) \wedge (Z_j = b) ] \geq \frac { ||x_{i,a}||}{\sqrt R} \cdot  \frac { ||x_{j,b}||}{\sqrt R} \geq \frac 1R \langle x_{i,a} , x_{j,b} \rangle \]
where we used Cauchy-Schwarz  again. Note that right-hand side is at least a $1/R$ fraction of the contribution of the constraint to the cost function, and so the rounding satisfies, on average, a number of constraints that is at least a $1/R$ fraction of the SDP value.

The above analysis is tight if all vectors have length $R^{-1/2}$ and if $x_{i,a}$ and $x_{j,b}$ are parallel, which means that in order to improve the $1/R$ approximation we need to devise a different rounding scheme.

A typical approach to round SDP formulations like \eqref{sdp} is to pick a random vector $g$ from a Gaussian distribution and assign to $X_i$ the value $a$ such that $\langle x_{i,a}, g\rangle$ is largest, or to pick a random value $a$ among those for which the inner product is larger than a certain threshold. The problem with this approach in our context is that if, say, $||x_{i,a}|| = R^{-.5}$ and $||x_{i,b}|| = R^{-.4}$, the assignment $X_i = b$ might be exponentially (in $R$) more likely than the assignment $X_i = a$, even though constraints that require $X_i = a$ give a contribution to the cost function that is polynomially (in $R$) related to the contribution of constraints that require $X_i = b$. This issue is related to the fact that  rounding algorithms such as the one of Charikar, Makarychev and Makarychev work only in instances in which a large fraction of constraints are satisfiable.

We develop a different type of rounding scheme, that we believe is new. For each vector $x_{i,a}$ we compute a probability $p_{i,a}$, proportional to the length $||x _{i,a}||$ which is a ``target probability'' with which we would like to have $X_i = a$. Then we compute a threshold $t_{i,a}$ such that, if $N$ is a standard normal distribution, 
\[ \Pr [ N \geq t_{i,a} ] = p_{i,a} \]
Finally, we pick a vector $g$ such that each coordinate of $g$ is a standard normal distribution, and, ideally, we would like to set $X_i = a$ if and only if
\[ \langle x_{i,a} , g \rangle \geq || x_{i,a} || \cdot t_{i,a} \]
an event that, indeed, happens with  probability $p_{i,a}$. Unfortunately, for the same $X_i$, we can have the above event hold for more than one value of $a$. We resolve this ambiguity by creating a short list $L_i$ of all the values $a$ for which the above event holds, and then by picking randomly from the short list.

This way we construct a random assignment $(Z_1,\ldots,Z_n)$ in which each $Z_i$ takes value $a$ with probability proportional to $|| x_{i,a}||$, and such that the events $Z_i = a$ and $Z_j = b$ are positively correlated if the angle between $x_{i,a}$ and $x_{j,a}$ is small. 

We describe our rounding algorithm more precisely in Section \ref{sec.rounding}. The analysis of the rounding algorithm relies on two key lemmas, showing, for each constraint $(X_i = a) \wedge (X_j = b)$, a lower bound to the probability that
$a$ ends up in the shortlist $L_i$ and $b$ ends up in the shortlist of $L_j$ (this is proved in Section \ref{sec.lower}) and an upper bound to the size of the shortlists (proved in Section~\ref{sec.upper}). Proving the two key lemmas requires several 
facts about the behavior of Gaussian vectors in certain conditional distributions, and we develop the required theory in Section \ref{sec.prelim}.

\section{The Rounding Algorithm}
\label{sec.rounding}

We start by assigning a probability
\begin{equation} p_{i,a} := \frac12\cdot\parenth{\frac { ||x_{i,a}|| }{\sqrt R } +\frac1R} \end{equation}
to each variable $X_i$ and each value $a$. This is similar to the probability $||x_{i,a}||/\sum_b ||x_{i,b}|| \geq ||x_{i,a}||/ \sqrt R$
in the naive rounding described in the introduction, but for technical reasons it is convenient to make sure that the probabilities are always at least $\Omega(1/R)$. Note also that the probability are always at most $1/\sqrt R$.

Then we define the  threshold $t_{i,a}$ as the number such that, if $N$ is  a standard normal distribution we have
\[ \Pr [ N \geq t_{i,a} ] = p_{i,a} \]
Note that $t_{i,a} = \Theta ( \sqrt {\log k })$.

Note also that for a Gaussian-distributed  vector $g$ (that is, a vector such that
each coordinate is a standard normal distribution) we have that
\[ \pr_r [ \langle g, x_{i,a} \rangle \geq ||x_{i,a}|| \cdot  t_{i,a} ] = p_{i,a}
.\] 
The rounding algorithm selects a 'short-list' $L_i$ of possible
assignments for each variable $i$, by picking one Gaussian vector $g$,
and defining for each $i$ the list $L_i$ by
\[ L_i := \{ a : \langle x_{i,a} , g \rangle \geq ||x_{i,a} || \cdot t_{i,a} \} . \]
The final assignment $Z_i$ for the $i$'th variable is chosen by
selecting an element of $L_i$ at random (no assignment is chosen if
$L_i$ is empty). 

\subsection{Analysis of the Algorithm}

Our analysis will proceed via the following two  results. 

\begin{lemma} \label{lower.bound} For every $i\neq j$ and every values $a,b$, 
\[ \Pr [ a\in L_i \wedge b \in L_j ] \geq \Omega(\log R) \cdot \frac 1R \cdot \langle x_{i,a},x_{j,b} \rangle \]
\end{lemma}

\begin{lemma} \label{upper.bound}  There is a constant $U$ such that
for every $i\neq j$ and every values $a,b$, 
\[ \Pr [ |L_i| \leq U \wedge |L_j| \leq U  \ | \ a\in L_i \wedge b\in L_j ] \geq \Omega(1) \]
\end{lemma}

We prove Lemma \ref{lower.bound} in Section \ref{sec.lower} and Lemma \ref{upper.bound} in Section \ref{sec.upper}. 
The approximation follows easily from them.

\begin{theorem}[Main] \label{th.main} The rounding algorithm achieves an $\Omega\left( \frac 1 R \log R \right)$ approximation ratio.
\end{theorem}

\begin{proof}
If we let $Z_1,\ldots,Z_n$ denote the randomized assignment created by the rounding algorithm, we want to show that the expected number of constraints satisfied by the assignment is at least an $\Omega(\log R /R)$ fraction of the optimum of the SDP, and hence at least an $\Omega(\log R / R)$ fraction of the optimum of the 2CSP-$R$ problem.

It is enough to show that, for every constraint $C$ of the form $(X_i = a) \wedge (X_j = b)$, the probability that the constraint is satisfied by the algorithm is at least an $\Omega(R^{-1}\log R)$ fraction the contribution of the constraint to the objective function, that is,
\begin{equation} \Pr [ Z_i = a \wedge Z_j = b ] \geq \Omega \left( \frac {\log R} R \right) \cdot \langle x_{i,a} , x_{j,b} \rangle \end{equation}
Which follows by observing that
\[ \Pr [ Z_i = a \wedge Z_j = b ] =  \Pr [ Z_i = a \wedge Z_j = b | a\in L_i \wedge b\in L_j ] \cdot \Pr [ a\in L_i \wedge b\in L_j ] \]
\[ \geq \Omega \left( \frac {\log R} R \right) \cdot \langle x_{i,a},x_{j,b} \rangle \cdot \Pr [ Z_i = a \wedge Z_j = b | a\in L_i \wedge b\in L_j ] \]
\[ \geq \Omega  \left( \frac {\log R} R \right) \cdot \langle x_{i,a},x_{j,b} \rangle \]
by Lemma \ref{lower.bound} and Lemma \ref{upper.bound}.\end{proof}

\section{Gaussian Estimates}
\label{sec.prelim}

In this section, we state and prove some facts about Gaussian Distributions which we use in the proofs that follow.

\begin{definition}[Normalized Gaussian vectors]
  A random variable $g\sim N(0,1)$ with normal distribution, zero expectation,
  and variance $1$ is called a {\em normalized Gaussian variable}. A
  normalized Gaussian vector in $\R^n$ is a random variable $g$
  with values in $\R^n$ whose coordinates are independent normalized
  Gaussians.
\end{definition}

The following simple claim is well known.
\begin{claim}
  If $g$ is a normalized Gaussian vector in $\R^n$ then $\inner{g,u}$
  is a normalized Gaussian variable for any unit vector $u\in\R^n$. 
\end{claim}

\begin{definition}[Gaussian vectors in a subpace]
Let $V\in\R^n$ be a subspace. A
  normalized Gaussian vector in $V$ is a random variable $g$
  with values in $V$ such that $\inner{g,v}$
  is a normalized Gaussian variable for any unit vector $v\in V$.
\end{definition}

It is easy to verify the following.

\begin{claim}\label{claim:sub-gaussian}
Let $V\in\R^n$ be a subspace, and let $g$ be 
  normalized Gaussian vector in $V$. Then for every unit vector
  $u\in\R^n$, $\inner{g,u}$ is a normal random variable with mean zero
  and variance at most $1$. 
\end{claim}

\begin{definition}
  We define the function $p:\R\to\R$ by $p(t)= \Pr\brac{g>t}$,
  where $g$ is a normalized Gaussian random variable. $p$ is the
  complement to $1$ of the cumulative distribution function for the
  Gaussian distribution.  
\end{definition}

The following is a well known bound on $p(t)$.

\begin{claim} (Gaussian Approximation) \label{claim_gaussianapprox}
Let $g$ be a standard normal random variable. Let $p(t) =\Pr[g>t]$. Then or any $t\in \mathbf{R}$, 
$$ \Omega\parenth{ \frac{e^{-t^2/2}}{t} }\leq\frac{t}{\sqrt{2\pi}(t^2+1)}e^{-t^2/2}\leq p(t) \leq \frac{1}{\sqrt{2\pi}t} e^{-t^2/2}$$
\end{claim}

The proof of the above claim can be found, for example, in \cite{CMM06}, Lemma A.1.

\begin{claim}\label{claim:ln-p-bound}
  There is a constant $t_0$ such that for every $t\geq t_0$, $\frac{t^2}{2}\leq\ln\parenth{\frac{1}{p(t)}} \leq t^2$
\end{claim}
\begin{proof}
  This follows immediately from Claim~\ref{claim_gaussianapprox}.
\end{proof}

We next present two key claims.

\begin{claim} (Lower Bound for Change in Probability)\label{claim_lowerchange}
 There is a constant $t_0$ such that for every $t\geq t_0$ and $\alpha\in[0,1]$
$$ p((1-\alpha)t)=\Pr[g>(1-\alpha)t]\geq \alpha\cdot p(t) \cdot \ln \left(\frac{1}{p(t)}\right)$$
\end{claim}
\begin{proof}
Let $D(\alpha)=p((1-\alpha)t)-\alpha p(t)\ln \frac{1}{p(t)}$. We will show that $D(\alpha)\geq 0$ for all $\alpha \in [0,1]$.

We observe that $D(0)\geq 0$. 
Using the fact that $p(t)=\frac{1}{\sqrt{2\pi}}\int_{t}^{\infty}e^{-r^2/2} dr$ and taking the derivative with respect to $\alpha$, we will show that the derivative in question is non-negative, thus the function is monotone increasing, which implies that $D(\alpha)\geq D(0)\geq 0$.

 \begin{align*}
\frac{d}{d\alpha}D(\alpha)=&\frac{d}{d\alpha}\left( \int_{(1-a)t}^{\infty} e^{-r^2/2}dr \right)-p(t)\ln \frac{1}{p(t)}\\
=&\frac{t}{\sqrt{2\pi}}e^{-(1-\alpha)^2t^2/2}-p(t)\ln \frac{1}{p(t)}\\
\geq& \frac{t}{\sqrt{2\pi}}e^{-t^2/2}e^{-\alpha^2t^2/2}e^{\alpha t^2}-\frac{1}{\sqrt{2\pi}t}e^{-t^2/2}\cdot \ln \left(\frac{\sqrt{2\pi}(t^2+1)e^{t^2/2}}{t}\right)\\
\geq & \frac{t}{\sqrt{2\pi}}e^{-t^2/2}e^{-\alpha^2t^2/2}e^{\alpha t^2}-\frac{t^2}{\sqrt{2\pi}t}e^{-t^2/2}\geq \frac{t}{\sqrt{2\pi}}e^{-t^2/2}e^{-\alpha^2t^2/2}e^{\alpha t^2}-\frac{t}{\sqrt{2\pi}}e^{-t^2/2}\\
=&\frac{t}{\sqrt{2\pi}}e^{-t^2/2}\left(e^{-\alpha^2t^2/2}e^{\alpha t^2}-1\right)\geq 0
\end{align*}

Here, the first inequality follows from claim \ref{claim_gaussianapprox} above, the second inequality follows from claim \ref{claim:ln-p-bound}
and the last inequality follows from observing that, for $\alpha\leq 1$, we have 
\[\alpha^2t^2/2 \leq \alpha t^2 \Rightarrow e^{\alpha^2t^2/2}\leq e^{\alpha t^2}\]

\end{proof}

\begin{claim}(Upper Bound for Change in Probability)\label{claim:upper-power}

  There is a constant $t_0$ such that for every $t\geq t_0$ and
  $\alpha\in[0,1]$, 
  \begin{equation}
    \label{eq:lower2}
    p\parenth{(1-\alpha)t}\leq p(t)^{\parenth{1-3\alpha}}.
  \end{equation}
\end{claim}
\begin{proof} 

Let $D(\alpha)= -p((1-\alpha)t)+ p(t)^{\parenth{1-3\alpha}}$. We will show that $D(\alpha)\geq 0$ for all $\alpha \in [0,1]$ and $t>t_0$.
We observe that $D(0)= 0$. Using the fact that $p(t)=\frac{1}{\sqrt{2\pi}}\int_{t}^{\infty}e^{-r^2/2} dr$ and taking the derivative of $D(\alpha)$ with respect to $\alpha$, we will show that the derivative in question is non-negative, thus the function is monotone increasing, which will imply that $D(\alpha)\geq D(0)= 0$ and will complete the proof .

 \begin{align*}
\frac{d}{d\alpha}D(\alpha)=&-\frac{d}{d\alpha}\left( \int_{(1-a)t}^{\infty} e^{-r^2/2}dr \right)+3p(t)^{(1-3\alpha)}\ln \frac{1}{p(t)}\\
\geq &-\frac{t}{\sqrt{2\pi}}e^{-(1-\alpha)^2t^2/2}+3 \frac{t^2}{2} p(t)^{(1-3\alpha)}\\
\geq &\frac{3}{2}t^2e^{-(1-3\alpha)t^2/2}\frac{1}{\sqrt{2\pi}(t+1)}-\frac{t}{\sqrt{2\pi}}e^{-(1-\alpha)^2t^2/2}\\
= & \frac{t}{\sqrt{2\pi}}e^{-t^2/2}\left(\frac{3t}{2(t+1)}e^{3\alpha t^2/2}-e^{\alpha t^2-\alpha^2t^2/2}\right)\\
\geq &\frac{t}{\sqrt{2\pi}}e^{-t^2/2}\left(e^{3\alpha t^2/2}-e^{\alpha t^2-\alpha^2t^2/2}\right) \geq 0
\end{align*}

here, the first inequality follows from claim \ref{claim:ln-p-bound}, the second inequality follows from claim \ref{claim_gaussianapprox} and the last inequality from the observation that \[e^{3\alpha t^2/2}\geq e^{\alpha t^2-\alpha^2t^2/2}\]

\end{proof}

\noindent The following claim is an easy corollary of  Claim~\ref{claim:upper-power}.

\begin{claim}\label{claim:threshold-advantage-corollary}
  For $t>1$ and $\alpha\leq\frac1{t^2}$,     
$p\parenth{(1-\alpha)t}\leq  O\parenth{p(t)}$. 
\end{claim}
\begin{proof}
  Using Claim~\ref{claim:upper-power}, it is enough to show that
  $p(t)^{-3\alpha}\leq O(1)$. But this is clear from the bound
  $p(t)\leq \frac{1}{\sqrt{2\pi}} \cdot \frac{e^{-t^2/2}}{t} $ and the
  bound on $\alpha$. 
\end{proof}

\begin{claim}\label{claim:wedge-lower-bound}
  Let $u,v\in \R^n$ be unit vectors such that $\inner{u,v}\geq 0$, and
  let $g$ be a normalized Gaussian vector in $\R^n$. Then for any
  $t_1,t_2>0$, $\Pr\brac{\inner{g,u}>t_1 \ve \inner{g,v}>t_2}\geq
  p(t_1)\cdot p(t_2)$.
\end{claim}
\begin{proof}
  The inner products $\inner{g,u}>t_1$ and $\inner{g,v}>$ are
  normalized Gaussian variables. The claim follows since because
  $\inner{u,v}\geq 0$, the correlation between these Gaussian variables
    is non-negative. We omit the simple
    details.\footnote{Alternatively, one can embed the quadrant
      $\set{x_1>t_1}\cap\set{x_2>t_2}$ into the event in question via
      a measure preserving map}
\end{proof}

\begin{claim}\label{claim:wedge-condition-bound}
 Let $u$ and $v$ be two unit vectors in $\R^2$ such that
 $\inner{u,v}\geq 0$, and let $g=(g_1,g_2)\in\R^2$ be a random vector whose
 coordinates are independent normalized Gaussians. Also let
 $t_1,t_2\gg 1 $ be thresholds such that $t_1\leq t_2\leq 4t_1$.  Then
 \begin{equation*}
   \Pr\brac{\max\set{g_1,g_2}>5t_1\;\big\vert \inner{g,u_1}\geq
     t_1\ve\inner{g,u_2}\geq t_2}\leq 1/2.
 \end{equation*}
\end{claim}
\begin{proof}
 Using Bayes' law we have
 \begin{align*}
     \Pr\brac{g_1>5t_1\;\big\vert \inner{g,u_1}\geq
     t_1\ve\inner{g,u_2}\geq t_2}&=\frac{  \Pr\brac{g_1>5t_1\;\ve \inner{g,u_1}\geq
     t_1\ve\inner{g,u_2}\geq t_2}}{\Pr\brac{\inner{g,u_1}\geq
     t_1\ve\inner{g,u_2}\geq t_2}} \\ &\leq
\frac {\Pr\brac{g_1>5t_1}}     { p(t_1)p(t_2) }  \qquad(\text{by
 Claim~\ref{claim:wedge-lower-bound}}) \quad \\ &\leq
\frac {p(5t_1)}     { p(4t_1)^2 } \leq \frac14\;,
 \end{align*}
 where the last inequality is by Claim~\ref{claim_gaussianapprox}. The same computation also bounds\\
 $\Pr\brac{g_2>5t_1\;\big\vert \inner{g,u_1}\geq
   t_1\ve\inner{g,u_2}\geq t_2}$ by $\frac14$, and we get the desired
 claim using the union bound.
\end{proof}

\noindent The following is an immediate corollary of
Claim~\ref{claim:wedge-condition-bound}.

\begin{claim}\label{claim:wedge-norm-bound}
  Let $u$ and $v$ be two unit vectors in $\R^2$ such that
  $\inner{u,v}\geq 0$, and let $g=(g_1,g_2)\in\R^2$ be a random vector whose
  coordinates are independent normalized Gaussians. Also let
  $t_1,t_2\gg 1 $ be thresholds such that $t_1\leq t_2\leq 2t_1$.  Then
  \begin{equation*}
    \Pr\brac{\norm{g}>10t_1\;\big\vert \inner{g,u_1}\geq
      t_1\ve\inner{g,u_2}\geq t_2}\leq 1/2.
  \end{equation*}
\end{claim}

\section{Proof of Lemma \protect{\ref{lower.bound}}}
\label{sec.lower}

In order to proof Lemma \ref{lower.bound} we need the following result.

\begin{lemma} \label{bound.condition} Suppose that $p_{j,b} \geq p_{i,a}$. Then 
\[ \Pr [ b \in L_j | a \in L_i ]\geq \Omega(cos(\theta)\cdot p_{j,b}\cdot \ln \left(\frac{1}{p_{j,b}}\right))\]
where $\theta =\widehat{x_{i,a},x_{j,b}} $ 
\end{lemma}
\begin{proof} 

  By definition, $p_{j,b} \geq p_{i,a}$ implies that $\|x_{j,b}\|\geq \|x_{i,a}\|$ and $t_{i,a}\geq t_{j,b}$, where $t_{i,a}\geq \Omega(\sqrt{\log \left (\frac{1}{p_{i,a}}\right )})$ and $t_{j,b}\geq \Omega(\sqrt{\log \left (\frac{1}{p_{j,b}}\right )})$ are chosen so that $$\Pr\brac{a\in L_i}=\Pr\brac{\langle \frac{x_{i,a}}{\|x_{i,a}\|},g \rangle \geq t_{i,a}}=p_{i,a}$$ and $$\Pr\brac{b\in L_j}=\Pr\brac{\langle \frac{x_{j,b}}{\|x_{j,b}\|},g \rangle \geq t_{j,b}}=p_{j,b}$$ Here $g$ is our random gaussian.

  Let $g = g_\parallel+g_\perp+g_0$ where
$g_\parallel$, is in the direction of $x_{i,a}$, $g_\perp$ is
in the direction perpendicular to $x_{i,a}$ in the plane spanned by $x_{i,a}$ and
$x_{j,b}$ and $g_0$ is perpendicular to the plane spanned by $x_{i,a}$ and $x_{j,b}$. Denote by $\theta$ the angle $\widehat{x_{i,a},x_{j,b}}$ and note that since, by the SDP constraints, $\langle x_{i,a},x_{j,b}\rangle \geq 0$ it holds that $cos(\theta) =  \langle x_{i,a} ,x_{j,b} \rangle \cdot \frac 1 {||x_{i,a}|| \cdot ||x_{j,b} ||}\geq 0$.
  
  \begin{align*}
  \label{eq:4}
  \Pr [ b \in L_j | a \in L_i ] = & \Pr\brac{\left. \langle \frac{x_{j,b}}{\|x_{j,b}\|},g \rangle > t_{j,b} \ \
  \right\vert \ \ \langle \frac{x_{i,a}}{\|x_{i,a}\|} ,g \rangle > t_{i,a}} \\
= & \Pr\brac{\left.  \langle \frac{x_{j,b}}{\|x_{j,b}\|},g_\parallel \rangle + \langle \frac{x_{j,b}}{\|x_{j,b}\|},g_\perp \rangle > t_{j,b} \ \
  \right\vert \ \ \langle \frac{x_{i,a}}{\|x_{i,a}\|} ,g_\parallel \rangle > t_{i,a}} \\
= & \Pr\brac{ \left \|g_\parallel\|cos(\theta)+ \|g_\perp\|sin(\theta)  > t_{j,b} \ \
  \right\vert \ \ \|g_\parallel\| > t_{i,a}} \\
  \geq & \Pr\brac{ \left. cos(\theta)t_{i,a}+ \|g_\perp\|sin(\theta)  > t_{j,b} \ \
  \right\vert \ \ \|g_\parallel\| > t_{i,a}} \\
  = & \Pr\brac{ cos(\theta)t_{i,a}+ \|g_\perp\| sin(\theta)  > t_{j,b}}\\
  \geq & \Pr\brac{ cos(\theta)t_{j,b}+ \|g_\perp\|sin(\theta)  > t_{j,b} }\\
   = & \Pr\brac{ \|g_\perp\|  > t_{j,b} \left(\frac{1-cos(\theta)}{sin(\theta)}\right) }\\
\end{align*}
  
  Where the last inequality follows from the fact that $t_{i,a}\geq t_{j,b}$ and $cos (\theta) \geq 0$.
  
  We now observe that
\begin{align*}
\frac{1-cos(\theta)}{sin(\theta)}=&\frac{\sqrt{(1-cos (\theta))}^2}{1-cos (\theta)^2}=\frac{\sqrt{(1-cos (\theta))}}{\sqrt{(1+cos (\theta))}}\\
\leq& \sqrt{1-cos(\theta)}\leq 1-\frac{1}{2}\cos (\theta )
\end{align*}

and we can write the expression above as 
\[ \Pr [ b \in L_j | a \in L_i ]\geq \Pr\brac{ \|g_\perp\|  > t_{j,b}\left(1-\frac{1}{2}cos(\theta)\right) }\]

Using claim \ref{claim_lowerchange} 
\[ \Pr [ b \in L_j | a \in L_i ]\geq \Omega(cos(\theta)\cdot p_{j,b}\cdot \ln \left(\frac{1}{p_{j,b}}\right))\]

 \end{proof}
 
 Now we are ready to prove Lemma \ref{lower.bound}.
 
\begin{proof}[Proof of Lemma \ref{lower.bound}] Without loss of generality we
can assume  $p_{j,b} \geq p_{i,a}$. Then 
\[ \Pr [ a\in L_i \wedge b \in L_j ]  = \Pr [ a\in L_i ] \cdot \Pr [ b \in L_j \ | \ a \in L_i  ] \] 
Using Lemma \ref{bound.condition}, the above is at least
\[p_{i,a}\cdot \Omega(cos(\theta)\cdot p_{j,b}\cdot \ln \left(\frac{1}{p_{j,b}}\right))\]
where $p_{i,a} \geq \frac 12 \frac {||x_{i,a}||}{\sqrt R}$, $p_{j,b} \geq \frac 12 \frac {||x_{j,b}||}{\sqrt R}$,  $\ln \left(\frac{1}{p_{j,b}}\right) =\Omega(\log R)$, and

\[ cos(\theta) = \frac { \langle x_{i,a}, x_{j,b} \rangle}{||x_{i,a} || \cdot || x_{j,b} || } \]

Putting everything together gives us
\[ \Pr [ a\in L_i \wedge b \in L_j ]  \geq \frac{\Omega(\log R)}{R} \cdot \langle x_{i,a}, x_{j,b} \rangle\]

which completes the proof.
\end{proof}

\section{Proof of Lemma \protect{\ref{upper.bound}}}
\label{sec.upper}

For the proof of Lemma~\ref{upper.bound} we need to bound the
  increase in probability that a Gaussian variable passes certain
  thresholds, when the thresholds are lowered. This is done in the
  next lemma.
\def\range{R}
\begin{lemma}\label{lemma:mutliple-threshold-advantage}
  Let $t_1,\ldots,t_\range$ be positive thresholds with values between
  $t_{\min}\gg1$ and $t_{\max}$, such that
  $\sum_{b=1}^\range p(t_b)\leq1$, and let $t_{\max}$ and $t_{\min}$ be
  the maximum and minimum of those thresholds respectively. Also, let
  $s_b$ be positive advantages, and set $s^2=\sum_{b=1}^\range
  s_b^2$. Then for any $\ell\in(0,1)$, 
  \begin{equation}
    \label{eq:6}
  \sum_{b=1}^\range p(t_b-s_b)\leq \frac{s^2}{
      (t_{\min})^2}\cdot O\parenth{
    \frac1\ell+\frac{p(t_{\min})\cdot\parenth{t_{\max}}^4 }{{p(t_{\max})}^{3\ell}} }.
  \end{equation}
\end{lemma}
\def\eqdef{=}
\begin{proof}
  For all $b$ denote $\alpha_b\eqdef\frac{s_b}{t_{\min}}$, so $s_b\leq
  \alpha_bt_b$ for all $b$, and also 
  \begin{equation}
    \label{eq:3}
 \sum_{b=1}^\range
  \alpha_b^2=\frac{s^2}{\parenth{t_{\min}}^2}.   
  \end{equation}
We thus have
\begin{equation}
  \label{eq:4}
   \sum_{b=1}^\range p(t_b-s_b) \leq   \sum_{b=1}^\range p((1-\alpha_b)t_b).
\end{equation}
To bound~\eqref{eq:4} we partition the summands on the right-hand side
according to the value of $\alpha_b$. 
\paragraph{Small $\alpha_b$'s.} Let $I_s\eqdef\set{b:\ \alpha_b\leq
  \frac1{(t_{\max})^2}}$. By
Claim~\ref{claim:threshold-advantage-corollary},
\begin{equation}
  \label{eq:5}
   \sum_{b\in I_s} p((1-\alpha_b)t_b)\leq O\parenth{\sum_{b\in
     I_s}p(t_b)} \leq O\parenth{\sum_{b=1}^\range p(t_b)}=O(1). 
\end{equation}

\paragraph{Large $\alpha_b$'s.} Let $I_\ell\eqdef\set{b:\
  \alpha_b>\ell}$. Since any probability is at most $1$,
\begin{equation}
  \label{eq:5a}
   \sum_{b\in I_\ell} p((1-\alpha_b)t_b)\leq \abs{I_\ell}\leq \frac{s^2}{\parenth{\ell\cdot t_{\min}}^2},
\end{equation}
where we have used  \eqref{eq:3} to bound $\abs{I_\ell}$. 

\paragraph{Medium $\alpha_b$'s.} Let $I_m\eqdef\set{b:\
  \frac{1}{(t_{\max})^2}\leq \alpha_b\leq \ell}$. By
Claim~\ref{claim:upper-power} we get 
\begin{align}
  \sum_{b\in I_m} p((1-\alpha_b)t_b)\leq& O\parenth{\sum_{b\in
      I_m}p(t_b)^{\parenth{1-3\alpha_b}}} \leq
  O\parenth{ \frac{1}{p(t_{\max})^{3\ell}} \sum_{b\in
      I_m}p(t_b)} \notag\\ &\leq O\parenth{
    \frac{p(t_{\min})}{p(t_{\max})^{3\ell}} \cdot \abs{I_m} }
  \leq O\parenth{ \frac{p(t_{\min})}{p(t_{\max})^{3\ell}} \cdot
    \frac{s^2\cdot\parenth{t_{\max}}^2}{\parenth{t_{\min}}^2}},\label{eq:hoopla}
\end{align}
where to bound $\abs{I_m}$ we used the fact that all $\alpha_b$'s in
it are at least $\frac1{(t_{\max})^2}$ together with~\eqref{eq:3}.

\medskip \noindent Combining the inequalities~\eqref{eq:5},
\eqref{eq:5a} and \eqref{eq:hoopla}, we have~\eqref{eq:6}. 
\end{proof}

We are now ready to prove Lemma~\ref{upper.bound}. 

\begin{proof}[Proof of Lemma~\ref{upper.bound}]
Let $g$ be the Gaussian vector used for the rounding algorithm, and
let us write it as $g=g_\parallel+g_\perp$, where
$g_\parallel\in\spn\set{x_{i,a},x_{i,b}}$ and
$g_\perp\in\spn\set{x_{i,a},x_{i,b}}^\perp$. Note that Lemma~\ref{upper.bound} follows easily from the following two
claims:
\begin{claim}\label{claim:dumb}
  \begin{equation}
    \label{eq:8}
    \Pr\brac{\norm{g_\parallel}>20\sqrt{\log \range}\;\big\vert a\in L_i \ve
      b\in L_j } \leq 1/2,
  \end{equation}
\noindent and
\end{claim}

\begin{claim}\label{claim:dumber}
  For any vector $h\in\spn\set{x_{i,a},x_{i,b}}$ such that
  $\norm{h}\leq 20\sqrt{\log \range}$, 
  \begin{equation}
    \label{eq:7}
    \E\brac{|L_i|+|L_j| \;\big\vert\ \ g_\parallel=h }\leq O(1)
  \end{equation}
\end{claim}
(the identity of $g_\parallel$ already determines whether or not
$a\in L_i$ or $b\in L_j$, and therefore once we condition on
$g_\parallel=h$ there is no need to also condition on those events). 
We present the proofs of those two claims in the appendix.
Using the above claims the proof is complete.
\end{proof}

\begin{proof}[Proof of Claim~\ref{claim:dumb}]
Let $t_{i,a}$ and $t_{j,b}$ be the thresholds used by the rounding
algorithm to determine if $a\in L_i$ and $b\in L_j$ respectively, and
assume without loss of generality that $t_{i,a}$ is the smaller of the
two. 
By the definition of the rounding algorithm, we know that the
thresholds $t_{i,a}$ and $t_{j,b}$, as well and other threshold $t$ used
by our rounding, satisfy $\frac1{2\range}\leq p(t)\leq\frac{1}{\sqrt
  \range}$. It then follows from Claim~\ref{claim_gaussianapprox} that
these thresholds satisfy \[\frac{\sqrt {\log \range}}{2}\leq t\leq
2\sqrt{\log \range}.\]
In particular, we have $t_{i,a}\leq t_{j,b}\leq 4 t_{i,a}$. 
 By the SDP constraints, we know that
  $\inner{x_{i,a},x_{j,b}}\geq 0$. We also know that $g_\parallel$ is a
  normalized Gaussian in the span of these two vectors. Hence by Claim~\ref{claim:wedge-norm-bound},
  \begin{equation*}
\begin{split}
  Pr\brac{\norm{g_\parallel}>20\sqrt{\log \range}\;\big\vert a\in L_i \ve
    b\in L_j } \\
  \leq Pr\brac{\norm{g_\parallel}>10 t_{i,a}\;\big\vert
    \inner{g_\parallel,\frac{x_{i,a}}{\norm{x_{i,a}}}}\geq t_{i,a} \ve
    \inner{g_\parallel,\frac{x_{j,b}}{\norm{x_{j,b}}}} >t_{j,b} } \leq \frac12
  \end{split}
\end{equation*}
\end{proof}

\begin{proof}[Proof of Claim~\ref{claim:dumber}]
We will show that     $\E\brac{|L_j| \;\big\vert\ \ g_\parallel=h }\leq
O(1)$, as the bound on the conditional expectation of $|L_i|$ is
identical. By linearity of expectation, 
\begin{align*}
\E\brac{|L_j| \;\big\vert\ \
  g_\parallel=h } &=\sum_{b=1}^\range \Pr\brac{b\in L_j \;\big\vert\ \
  g_\parallel=h} = \sum_{b=1}^\range
\Pr\brac{\inner{g,\frac{x_{j,b}}{\norm{x_{j,b}}}} >t_{j,b} \;\big\vert\ \
  g_\parallel=h} \\ &= \sum_{b=1}^\range
\Pr\brac{\inner{g_\perp,\frac{x_{j,b}}{\norm{x_{j,b}}}} + \inner{h,\frac{x_{j,b}}{\norm{x_{j,b}}}} >t_{j,b}}.
\end{align*}
Since $g_\perp$ is a normalized Gaussian vector in a subspace,
Claim~\ref{claim:sub-gaussian} implies that $\inner{g_\perp,\frac{x_{j,b}}{\norm{x_{j,b}}}}$ has a normal
distribution with zero mean and variance bounded by $1$.  We therefore
have that, denoting $s_b\eqdef
\inner{h,\frac{x_{j,b}}{\norm{x_{j,b}}}}$, 
\begin{equation}\label{eq:almost-done}
  E\brac{|L_j| \;\big\vert\ \
  g_\parallel=h }\leq \sum_{b=1}^\range
p\parenth{t_{j,b}-s_b}.
\end{equation}
We woud like to now bound this sum using
Lemma~\ref{lemma:mutliple-threshold-advantage}. As we already noticed
in the proof of Claim~\ref{claim:dumb}, all the $t_{j,b}$'s are in the
range between $t_{\min}\geq \frac{\sqrt {\log \range}}{2}$ and
$t_{\max}\leq 2\sqrt{\log \range}$, such that $p(t_{min})\leq\frac1{\sqrt
  R}$ and $p(t_{\max})\geq\frac1{2R}$. It also follows from the SDP
relaxation constraints that the vectors $x_{j,b}$ are orthogonal, and therefore
$s^2=\sum_{b=1}^\range{s_b^2}\leq\norm{h}^2\leq 400\log \range.$ Finally,
the choice of the thresholds $t_{j,b}$ of the rounding algorithm
ensures that $\sum_{b=1}^\range p(t_{j,b})=1$. We can thus use the
parameter $\ell=.1$ in Lemma~\ref{lemma:mutliple-threshold-advantage},
and get 
\begin{align*}
  \eqref{eq:almost-done}\leq &\frac{s^2}{
      (t_{\min})^2}\cdot O\parenth{
    \frac1\ell+\frac{p(t_{\min})\cdot\parenth{t_{\max}}^4
    }{{p(t_{\max})}^{2\ell}} } \\ \leq&  O\parenth{
    1+\frac{\range^{-1/2}\cdot (\log \range)^2}{\range^{-0.3}}  } =O(1),
\end{align*}
as we wanted.

\end{proof}


\bibliography{2csp}
\section{Appendix}

\subsection{Integrality gap}

Khot and Vishnoi \cite{KV05} describe an integrality gap instance for a semidefinite programming 
relaxation of Unique Games. Our relaxation, applied to Unique Games, is the same as the one studied
by Khot and Vishnoi. We will refer to the full version of \cite{KV05}, which is available online
as a preprint \cite{KV13}. We need to slightly change the analysis of the ``completeness'' step, 
which is not optimized for the case of highly unsatisfiable instances in \cite{KV13}.

In the following, we refer to the {\em cost} of an assignment or of an SDP solution as
the number of satisfied constraints (respectively, the value of the objective function) divided
by the total number of constraints.

The construction of \cite{KV13} is parameterized by a value $\eta$, that we will set to $\frac 12 - \frac {1}{4\log R}$. In the proof of \cite[Lemma 3.6]{KV13}, Khot and Vishnoi prove that every solution
satisfies at most a
\[ R^{1- 2/(2-2\eta) } = R^{-1 + (2-4\eta)/(2-2\eta)} \]
\noindent fraction of constraints, which, by our choice of $\eta$, is at most 
\[ R^{-1} \cdot R^{2-4\eta} =  R^{-1} \cdot R^{1/\log R}   = \frac e{R} \]

Then they describe a feasible solution for the SDP relaxation and prove that it has cost at least $1-9\eta$. 
We need to argue the stronger fact that their solution has cost at least $(1-2\eta)^2$.
Khot and Vishnoi first describe a solution that satisfies all constraints except the $\langle x_{i,a} , x_{j,b} \rangle \geq 0$ nonnegativity constraints, and then obtain a feasible solution by taking a tensor product of each vector
with itself.

Before taking the tensor product, one can see that the cost of the solution is $\frac 1 R$ times the
expectation of the inner product $\langle x,y \rangle$ where $x,y$ are in $\{ -1,1\}^R$, $x$  is uniformly
distributed, and $y$ is obtained from $x$ by, independently for each coordinate $i$, setting $y_i = x_i$ with probability
$1-\eta$ and $y_i = -x_i$ with probability $\eta$. The expectation is clearly $R\cdot (1-2\eta)$ and so the cost
of the solution is $1-2\eta$. After taking the tensor products, the cost of the solution is $\frac 1R$ times the expectation of $(\langle x,y\rangle)^2$, where $x,y$ are distributed as above. Here we use the fact that $\E X^2 \geq (\E X)^2$,
and so the cost of the solution is at least $(1-2\eta)^2$.

Since we set $\eta = \frac 12 - \frac {1}{4\log R}$, the cost of the solution is $1/4(\log R)^2$.

In conclusion, we have an integrality gap instance in which the true  optimum is at most $e/R$ and the
optimum of the SDP relaxation is at lest $1/4(\log R)^2$, for an integrality gap of $\Omega ( R/(\log R)^2)$.

\subsection{UGC Hardness that Matches our Algorithm}

In this section, we show that the following 
\begin{claim} \label{claim_UGChardness}
For every $R\geq 2$, every $0 < \rho < 1$
and every $\epsilon >0$, it is UGC-hard to distinguish instances of Max 2LIN-$R$ in which at least a $\Omega(1/\log R)$
fraction of constraints are satisfiable, from instances where at most a $O(1\R)$ fraction
of constraints are satisfiable.
\end{claim}

We use the following result from Khot et al. \cite[Theorem 9, Section 6.3]{KKMO07}, who show that for every $R\geq 2$, every $0 < \rho < 1$
and every $\epsilon >0$ it is UGC-hard to distinguish instances of Max 2LIN-$R$ in which at least a $\rho + R^{-1} (1-\rho) -\epsilon$
fraction of constraints are satisfiable, from instances where at most a $R \cdot \Lambda_\rho (R^{-1}) +\epsilon$ fraction
of constraints are satisfiable; in the above expression $\Lambda_\rho (x)$ is a function that is defined in
 \cite[Definition 8, Section 4.2]{KKMO07}.

In the same paper,  they give the following bound on $\Lambda_\rho (x)$, which we repeat here for completeness.

\begin{corollary}(\cite[Proposition 1, Section 6.3]{KKMO07})
Let $\phi(x) =\frac{1}{\sqrt{2\pi}}e^{-x^2/2}$, the Gaussian density function. Let $t=t(x)$ be such that $p(t)=x$, where, as in Claim \ref{claim_gaussianapprox}, $p(t)=\int_{t}^{\infty}\phi(x)dx$ is the Gaussian tail probability function. Then for any $x\in[0,1/2)$ and any $\rho=\rho(x)$, $\rho \in [0,1]$ the following holds: $$\Lambda_\rho(x)\leq (1+\rho)\frac{\phi(t)}{t}p(t\sqrt{\frac{1-\rho}{1+\rho}})$$
\end{corollary}

\begin{proof}(Of Claim \ref{claim_UGChardness})
In order to prove claim \ref{claim_UGChardness}, we set $\rho = 1/12\log R$, and $x=1/R$. Using the estimates from claim \ref{claim_gaussianapprox} we obtain
$\frac{\phi(t)}{t} \leq O(p(t))$, which, from the corollary above, implies $$\Lambda_{1/12\log R}(1/R)\leq O(p(t))\cdot p(t\sqrt{\frac{1-1/12\log R}{1+1/12\log R}})$$
Using Claim \ref{claim:upper-power}, we have the upperbound: $p(t\sqrt{\frac{1-\rho}{1+\rho}}) \leq p(t\sqrt{1-2\rho}) \leq p(t(1-4\rho))\leq p(t)^{\parenth{1-12\rho}} $. We conclude that  $$\Lambda_{1/12\log R}(1/R)\leq O(p(t))\cdot p(t)^{\parenth{1-1/\log R}}\leq O(1/R^2 \cdot R^{1/\log R}) =O(1/R^2)$$

By using \cite[Theorem 9, Section 6.3]{KKMO07}, we obtain that it is UGC-hard to distinguish between the case where at least $\Omega(1/\log R)$ fraction of constraints are satisfiable and the case where at most $R \cdot O(1/R^2)=O(1/R)$ fraction of the constraints are satisfiable, which concludes the proof of the claim. 
\end{proof}
\end{document}